\documentclass[12pt]{nature}
\usepackage{graphicx}
\usepackage{amssymb, amsmath}
\usepackage{color}
\usepackage{hyperref}
\makeatletter
\let\saved@includegraphics\includegraphics
\AtBeginDocument{\let\includegraphics\saved@includegraphics}
\renewenvironment*{figure}{\@float{figure}}{\end@float}
\makeatother

\newcommand{\beam}{$\theta_{\mbox{\scriptsize maj}}\times\theta_{\mbox{\scriptsize min}}$}
\newcommand{\rah}{$^{\mbox{\scriptsize h}}$}
\newcommand{\ram}{$^{\mbox{\scriptsize m}}$}
\newcommand{\ras}{$^{\mbox{\scriptsize s}}$}
\newcommand{\decd}{$^{\circ}$}
\newcommand{\decm}{$'$}
\newcommand{\decs}{$''$}

\usepackage{nature}

\title{A likely flyby of binary protostar Z~CMa caught in action}

\author{Ruobing Dong$^{1}$, Hauyu Baobab Liu$^2$, Nicol\'as Cuello$^3$, Christophe Pinte$^4$, P\'eter \'Abrah\'am$^{5,17}$, Eduard Vorobyov$^{6,16}$, Jun Hashimoto$^{7}$, \'Agnes K\'osp\'al$^{5,13,17}$, Eugene Chiang$^8$, Michihiro Takami$^2$, Lei Chen$^5$, Michael Dunham$^9$, Misato Fukagawa$^{10}$, Joel Green$^{11}$, Yasuhiro Hasegawa$^{12}$, Thomas Henning$^{13}$, Yaroslav Pavlyuchenkov$^{14}$, Tae-Soo Pyo$^{15}$, Motohide Tamura$^{7,10,18}$}

\begin{document}

\maketitle

\begin{affiliations}
 \item Department of Physics \& Astronomy, University of Victoria, Victoria, BC, V8P 1A1, Canada
 \item Institute of Astronomy and Astrophysics, Academia Sinica, Taipei 10617, Republic of China
 \item Univ. Grenoble Alpes, CNRS, IPAG, F-38000 Grenoble, France 
 \item Monash Centre for Astrophysics (MoCA) and School of Physics and Astronomy, Monash University, Clayton VIC 3800, Australia
 \item Konkoly Observatory, Research Centre for Astronomy and Earth Sciences, E\"otv\"os Lor\'and Research Network (ELKH), Konkoly-Thege Mikl\'os \'ut 15-17, 1121 Budapest, Hungary
 \item University of Vienna, Department of Astrophysics, Vienna, 1180, Austria
 \item Astrobiology Center, National Institutes of Natural Sciences, 2-21-1 Osawa, Mitaka, Tokyo 181-8588, Japan
 \item Department of Astronomy, University of California at Berkeley, Berkeley, CA 94720-3411, USA
 \item Department of Physics, State University of New York at Fredonia, Fredonia, NY 14063, USA
 \item National Astronomical Observatory of Japan, 2-21-1, Osawa, Mitaka, Tokyo, 181-8588, Japan
 \item Space Telescope Science Institute (STScI), Baltimore, MD 21218, USA
 \item Jet Propulsion Laboratory, California Institute of Technology, Pasadena, CA 91109, USA
 \item Max-Planck-Institut f\"ur Astronomie, Heidelberg 69117, Germany
 \item Institute of Astronomy, Russian Academy of Sciences, Moscow, Russia
 \item Subaru Telescope, National Astronomical Observatory of Japan, National Institutes of Natural Sciences, HI 96720, USA
 \item Research Institute of Physics, Southern Federal University, Rostov-on-Don 344090, Russia
 \item ELTE E\"otv\"os Lor\'and University, Institute of Physics, P\'azm\'any P\'eter s\'et\'any 1/A, 1117 Budapest, Hungary
 \item Department of Astronomy, The University of Tokyo, Tokyo 113-0033, Japan
\end{affiliations}

\small

\begin{abstract}
Close encounters between young stellar objects in star forming clusters are expected to dramatically perturb circumstellar disks. Such events are  witnessed in numerical simulations of star formation\cite{offner08, bate12, kuruwita20}, but few direct observations of ongoing encounters have been made. Here we report sub-$0''.1$ resolution Atacama Large Millimeter Array (ALMA) and Jansky Very Large Array (JVLA) observations towards the million year old binary protostar Z~CMa in dust continuum and molecular line emission. A point source $\sim$4700 au from the binary has been discovered at both millimeter and centimeter wavelengths. It is located along the extension of a $\sim$2000 au streamer structure previously found in scattered light imaging, whose counterpart in dust and gas emission is also newly identified. Comparison with simulations shows signposts of a rare flyby event in action. 
Z~CMa is a ``double burster", as both binary components undergo accretion outbursts\cite{audard14}, which may be facilitated by perturbations to the host disk by flybys\cite{pfalzner08simulation, cuello19hydro, dullemond19,vorobyov21}.
\end{abstract}

Z Canis Majoris (hereafter Z~CMa) is a pre-main sequence binary system located at $\sim$1125 pc in the Canis Major R1 association (Methods). 
The southeast component belongs to a class of outbursting protostars called FUors, inferred based on the characteristic broad doubled optical absorption lines\cite{hartmann89}. The 0$''$.1 apart northwest component
has displayed repeated outbursts in the past few decades, resembling the EXor type outbursting protostars\cite{bonnefoy17}. 
The pair has been spatially resolved in adaptive optics observations at multiple epochs\cite{hinkley13, bonnefoy17}, and their motion is consistent with a face-on circular orbit composed of a 5$M_\odot$ (EXor component) and 1$M_\odot$ star (FUor component)\cite{millan02, canovas15zcma}.

\begin{figure}
\includegraphics[width=\textwidth,angle=0]{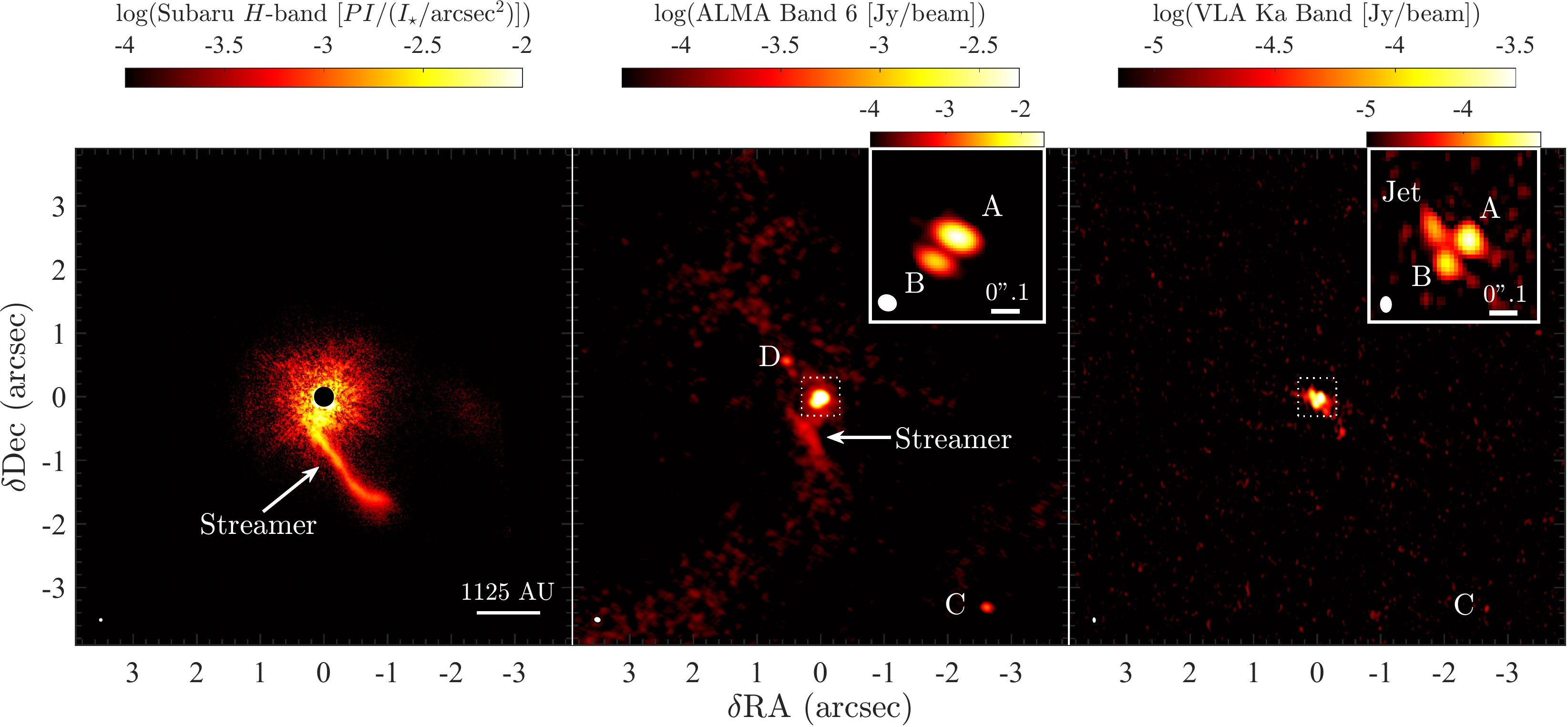}
\caption{{\bf Observations of the Z~CMa system.} The center is at component A (ALMA coordinates RA, Dec $=$ 07\rah03\ram43\ras.154,\  -11\decd33\decm06\decs.13). Left: Subaru $H$-band scattered light image with an angular resolution of 0.07 arcsec\cite{liu16subaru}. The polarized intensity (PI) is normalized to the stellar intensity flux ($I_\star$). Middle: ALMA 1.3 mm dust continuum emission with Briggs robust $=2$ (beam size 115$\times$95 mas, position angle = P.A. = 74$^\circ$). The inset is a zoom in of the inner 0.6$\times$0.6 arcsec region (dotted box) with Briggs Robust $=-2$ (beam size 71$\times$42 mas, P.A. = 68$^\circ$). Right: JVLA 9 mm dust continuum emission with Briggs Robust $=2$ (beam size 100$\times$66 mas, P.A. = 4$^\circ$). The inset is a zoom in of the inner 0.6$\times$0.6 arcsec region (dotted box) with Briggs Robust $=-2$ (beam size 62$\times$45 mas, P.A. = 1$^\circ$). In all panels, the angular resolution or the beam is marked at the lower left corner. The four point sources detected in the ALMA observations, and the counterparts of A, B, and C and a jet detected in the JVLA observations, are labeled. A version of the figure without annotations can be found in Extended data Fig.~\ref{fig:basic_noannotation}.}
\label{fig:basic}
\end{figure}

Previous spatially unresolved millimeter observations reported a total flux density of 26$\pm1$ mJy at 1.4 mm and 2.1$\pm0.1$ mJy at 7 mm\cite{alonsoalbi09}. 
We observed the Z~CMa system in continuum and molecular line emission at around 224 GHz (1.3 mm) using the Atacama Large Millimeter Array (ALMA; Methods), and in continuum emission at 33 GHz (9 mm) using the Jansky Very Large Array (JVLA; Methods). Fig.~\ref{fig:basic} shows the synthesized 1.3 and 9 mm continuum emission images. In total we detect four point like sources (A--D) at 1.3 mm. Two of them, A and B, are the central binary. Gaussian fitting of the flux indicates that both are surrounded by disks $\sim$20 au in size, potentially containing mm-sized dust particles. 
This is in line with millimeter observations of outbursting protostars in general\cite{liu16exor, liu17fuori}, 
which often show compact dust emission. At 9 mm, both sources are detected too, as well as a jet-like structure $0''.1$ in length likely originated from B. Gas observations show prominent outflow and cavity structures on large scales (Extended Data Figs.~\ref{fig:gas_big} and \ref{fig:gas_mom1_big}).

\begin{figure}
\includegraphics[width=0.9\textwidth,angle=0]{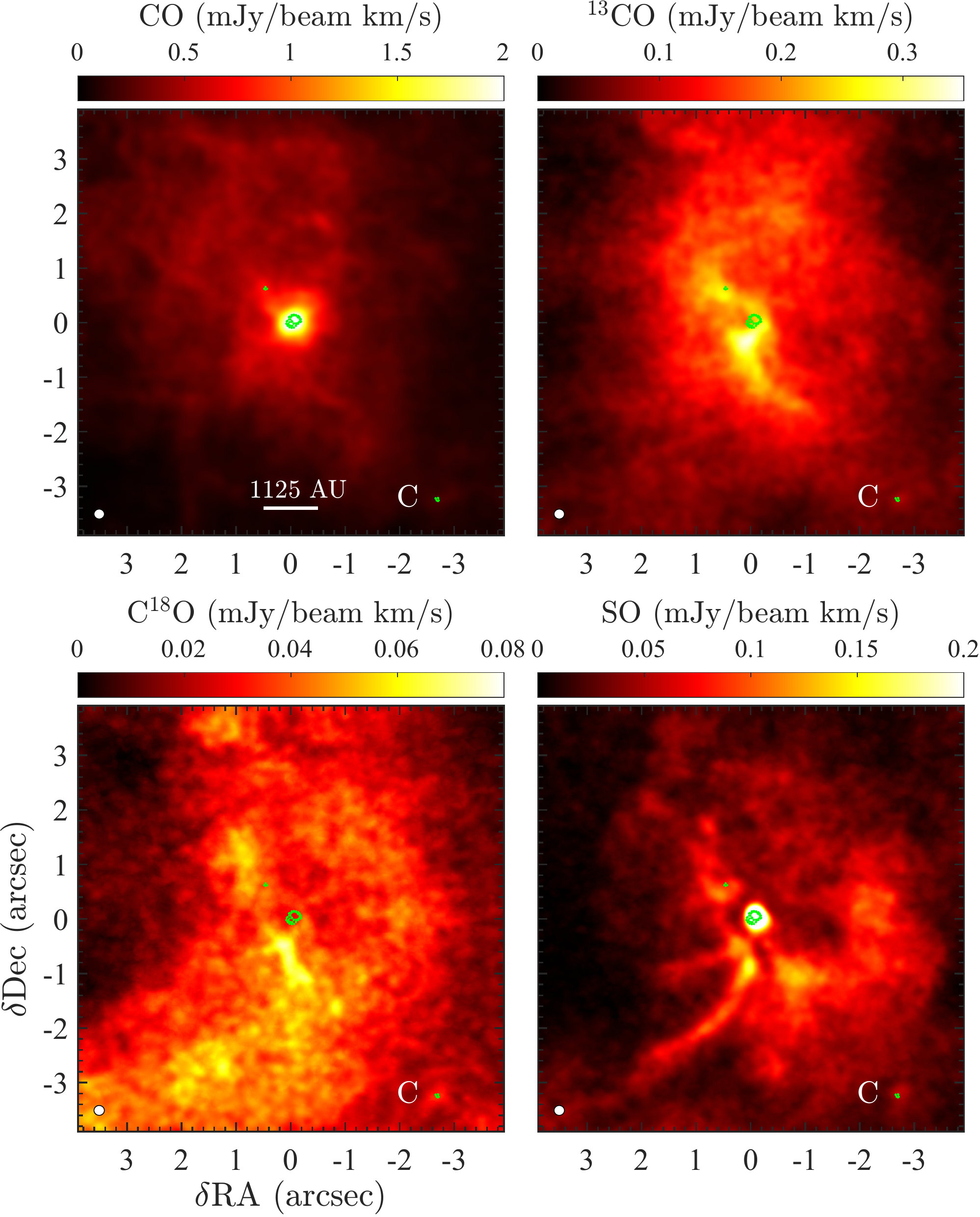}
\caption{{\bf Integrated intensity (moment 0) maps of the $^{12}$CO (2-1), $^{13}$CO (2-1), C$^{18}$O (2-1), and SO $^{3}\Sigma$ 6(5)-5(4) transitions.} The beam size ($0.19''\times0.18''$; P.A.=88$^\circ$) is marked at the lower left corner. The Robust $=0$ weighted 224 GHz continuum image (beam size $0''.075\times0''.047$; P.A.=65$^{\circ}$) is shown in green contours at 0.13 and 1.3 mJy beam$^{-1}$ levels (5 and 50 $\times$ the root mean square noises) to highlight the four compact continuum sources. The ALMA/JVLA source C is labeled.}
\label{fig:gas_small}
\end{figure}

The other two 1.3 mm sources, C and D, were previously unknown. In the continuum map imaged with Robust $=0$, C is detected with a flux density of $700\pm40$ mJy and a signal-to-noise ratio (SNR) of 27 at $4''.2$ (4725 au), while D is detected with a flux density of $550\pm24$ mJy and a SNR of 21 at $0''.8$ (900 au; Methods; Extended Data Table~\ref{tab:fluxes}).
They have spectral line counterparts, as shown in Fig.~\ref{fig:gas_small}
(most noticeable in C$^{18}$O and SO emission; Methods), that appear to connect to the 
circumbinary environment of Z~CMa. We conclude that both are likely physically associated with the system rather than being foreground or background sources. 
In addition, component C is detected at 9 mm by JVLA with SNR$=$5 (Methods). Gaussian fitting of its flux indicates that C is marginally resolved at both 1.3 and 9 mm with a deconvolved size of 40-50 au. Its spectral index $\alpha$ between the two wavelengths is 1.6$\pm$0.1. This low index may be explained by non-thermal contributions at 9 mm, a hot central source embedded in an optically thick medium\cite{li17}, or 
strongly scattering dust at 
mm wavelengths\cite{liu19}. Component D cannot be recovered in the JVLA observations. It may be a fragment that has recently formed from turbulent fragmentation in the disk or surrounding cloud\cite{kuffmeier19, lee19}, in which scenario it is only visible at mm but not cm wavelength due to its youth and associated insufficient dust growth. Further characterizations of these sources are presented in the Methods section and Supplementary Information, including the intensity weighted velocity (moment 1) maps. No other point sources have been detected in 1.3 mm continuum emission imaged with Robust $=0$ with a 3-$\sigma$ detection limit of 78 $\mu$Jy, equivalent to a dust mass of 3 $M_\oplus$ (Methods).

\begin{figure}
\includegraphics[width=\textwidth,angle=0]{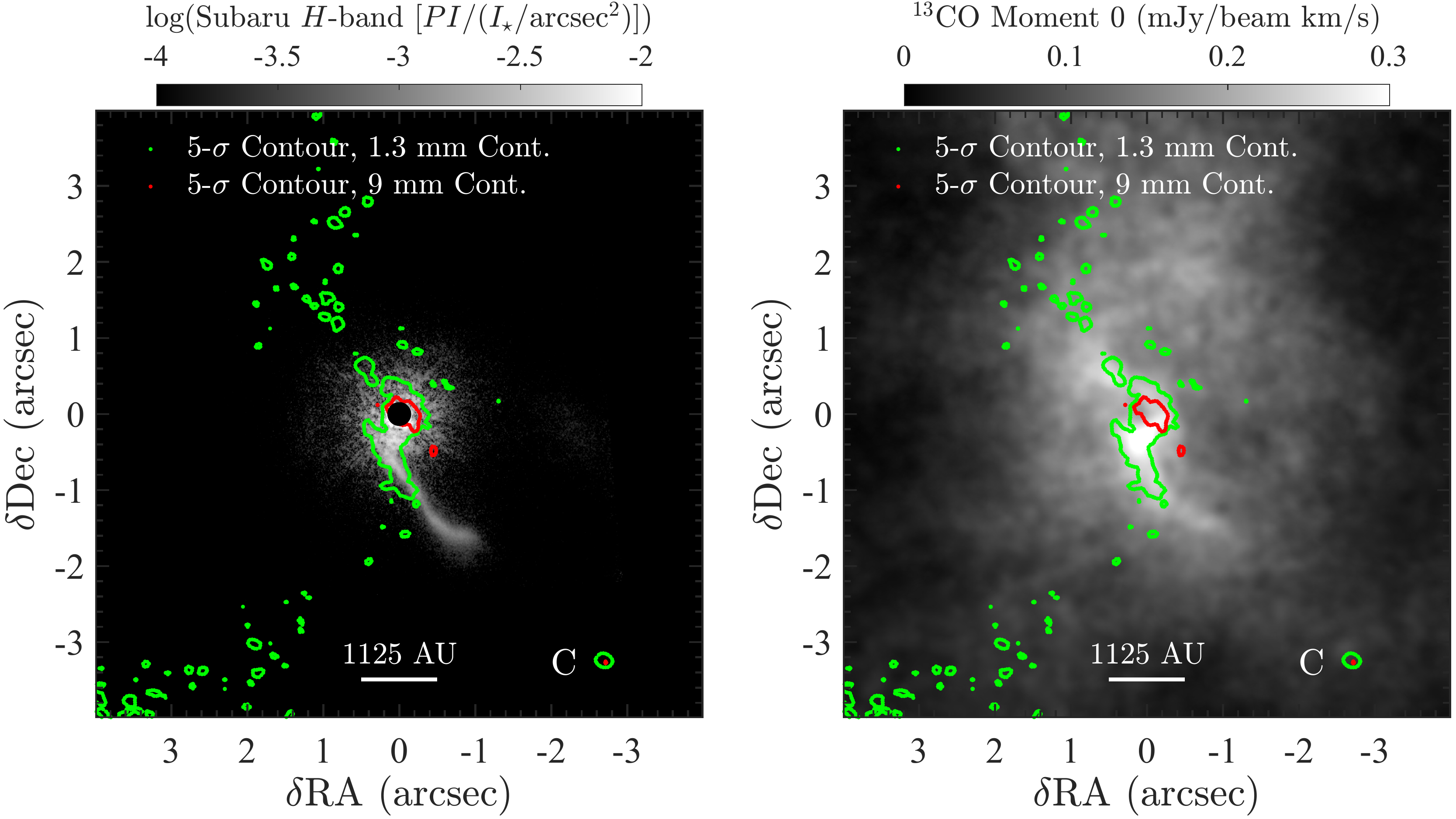}
\caption{{\bf Subaru $H$-band image (left) and $^{13}$CO moment 0 map (right) overlaid with the 5-$\sigma$ contours in ALMA 1.3 mm continuum emission (green) and in JVLA 9 mm continuum emission (red).} Both the ALMA and JVLA maps are imaged with Robust $=2$. The ALMA/JVLA source C is labeled.}
\label{fig:contour}
\end{figure}

Near-infrared imaging observations have revealed a large disk in scattered light extending to $\sim$1500 au\cite{liu16subaru}. Remarkably, the disk has a $\sim 2''$ long filamentary structure in the southwest, discovered in Keck adaptive optics observations\cite{millan02} and confirmed by VLT and Subaru in polarized light imaging\cite{canovas15zcma, liu16subaru} (Fig.~\ref{fig:basic}). In our observations the inner $\sim1''$ of this streamer is detected in 1.3 mm continuum, with a total flux density of 3$\pm0.6$ mJy and an estimated solid mass of 119$\pm24$ $M_\oplus$ (Methods). It is detected in $^{13}$CO and C$^{18}$O emission as well (Fig.~\ref{fig:contour}). The detection in 
dust and gas emission suggests that the streamer is a dust surface density enhancement.

The origin of the streamer has been a long-standing puzzle. It has been variously hypothesized to be the wall of a cavity created by a molecular outflow\cite{canovas15zcma}, a spiral arm produced by gravitational instability\cite{liu16subaru, dong16protostellar}, or the relic of a recently captured low mass ($\sim0.01M_\odot$) cloudlet\cite{dullemond19}.
The newly discovered source C at the end of the streamer points to another possible scenario that we consider the most likely: C may be a young stellar object (YSO) that has recently performed a flyby around Z~CMa, and the streamer is a structure produced by this tidal interaction event.

Tidal interactions between a protoplanetary disk and a flyby intruder are known to strongly perturb the disk\cite{pfalzner03}. Depending on the angle $\beta$ between the angular momentum vector of the disk and that of the intruder (with respect to the host), flyby events can be categorized as prograde ($\beta<90^\circ$) or retrograde ($90^\circ<\beta\leq180^\circ$). Prograde flybys typically induce a pair of spiral arms in the disk on global scale, with the primary arm pointing towards the intruder and the secondary arm roughly $180^\circ$ away in the azimuth. These arms are due to the combined effect of the intruder's gravitational perturbation and the displacement of the centre of mass of the system\cite{pfalzner03}. They are most prominent and symmetric when $\beta=0^\circ$ (coplanar and prograde flyby), and their strength and symmetry decrease as $\beta$ increases. Retrograde flybys have much weaker effects on the disk, unless it is deeply penetrating\cite{cuello19hydro}. In non-coplanar flybys, the primary arm can be lifted above the disk plane to keep pointing to the intruder, while the secondary arm is embedded in the disk\cite{cuello19hydro}.

\begin{figure}
\includegraphics[width=\textwidth,angle=0]{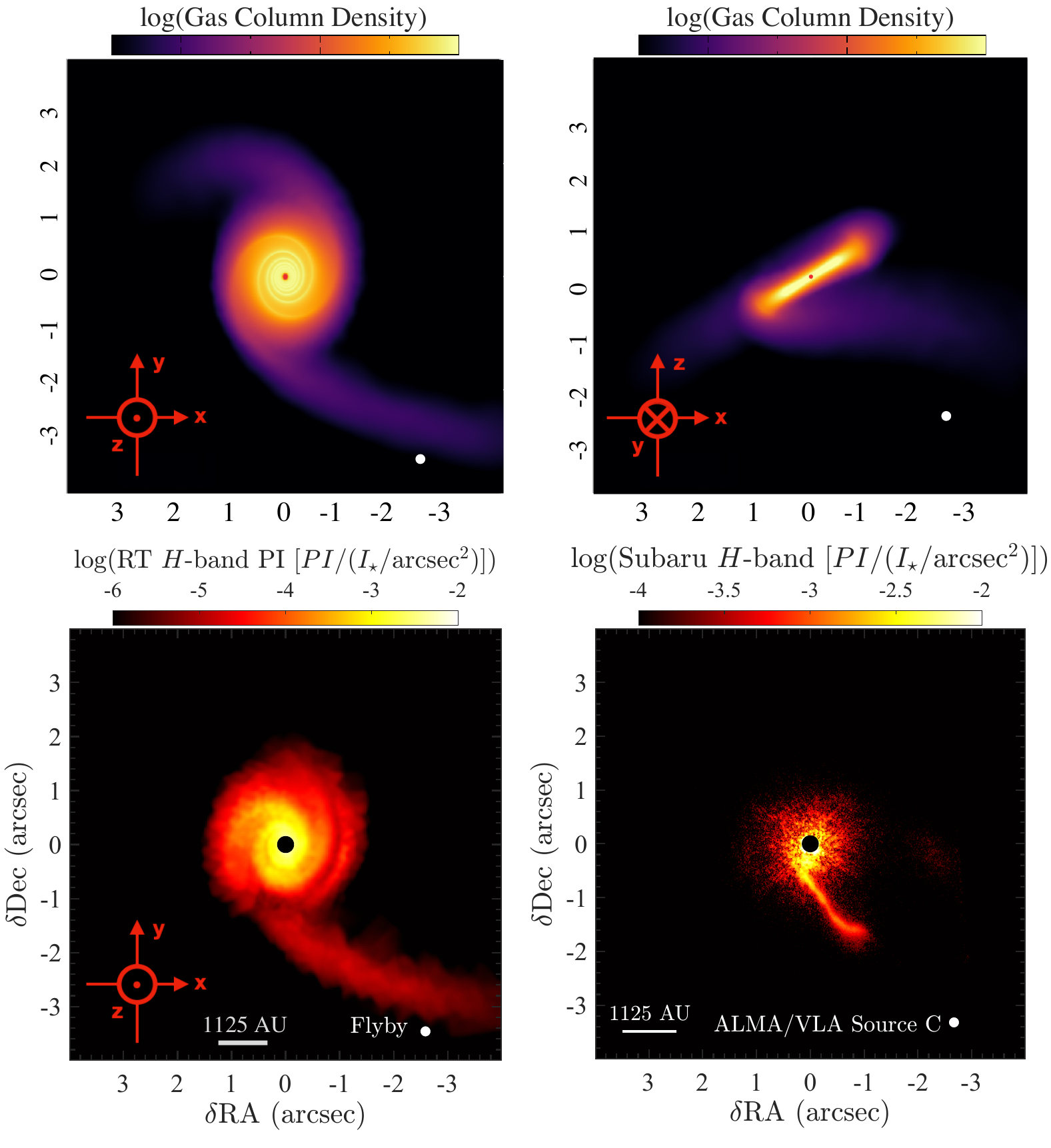}
\caption{{\bf Comparing simulations with observations.} (Top row) Two views of the gas column density in the hydrodynamical simulation of a stellar flyby. The system is shown $14$ kyr after the perturber has gone through pericentre. The resulting $H$-band synthetic image in polarized light obtained through radiative transfer simulation (bottom left) is to be compared with the Subaru $H$-band polarized light observation (bottom right). In the synthetic image the line of sight is in the $z$ direction.
}
\label{fig:simulation}
\end{figure}

Fig.~\ref{fig:simulation} shows a simulated flyby event (Methods; Extended Data Fig.~\ref{fig:sketch}), in which an intruder flies by a single disk-hosting primary with an intruder/primary mass ratio $q=0.3$. We do not expect the binary nature of Z~CMa to significantly affect the outcome of the flyby at distances more than $10$ times the binary separation. 
The intruder approaches the host on a prograde parabolic orbit inclined with respect to the host's disk by $\beta=45^\circ$. The pericenter is at $r_{\rm peri}=3000$~au and in the disk plane. The size of the initial disk surrounding the host is $R_{\rm out}=840$~au, thus the flyby is non-penetrating. 
A pair of spiral arms are produced in the density space with comparable strengths. The two arms, however, have different illuminations, because they are in different planes. The primary arm pointing to the flyby
is in the orbital plane of the flyby, and the light path from the host star is unobscured (the host star is assumed to be much more luminous than the intruder at near-infrared wavelengths). The arm is expected to be present in dust and gas emission due to its enhanced density\cite{cuello20flyby}. In contrast, the secondary arm is in the plane of the circumprimary disk and is much less well illuminated by the host star\cite{cuello20flyby}. 
The overall morphology of the model, with one prominent arm extending beyond the main disk and pointing towards the intruder, resembles the observations of the Z~CMa system (Fig.~\ref{fig:simulation}).
We note that the model does not represent a unique solution; instead it is one representative example that produces an overall match to the observations (Methods).

Component C has not been identified at optical and infrared wavelengths in the SDSS, Gaia, 2MASS, and WISE databases.
Analysis of archival Keck adaptive optics imaging observations of Z~CMa yields an upper limit of 21.9 mag for C in $H$-band apparent magnitude (Methods; Extended Data Fig.~\ref{fig:keck}), corresponding to the photosphere of a Myr-old planetary mass object at 1125 pc, or a solar mass YSO experiencing $\sim$10 mag of extinction\cite{baraffe15}. Past simulations and our tests indicate that an intruder with $q\lesssim0.1$ has difficulties in exciting the observed streamer, putting a lower limit of $\sim0.6M_\odot$ in mass for C (Methods). Therefore, C is likely heavily extincted. A flyby intruder may capture material from the host disk to form an envelope or a disk, depending on the angle and the level of penetration of the flyby\cite{cuello19hydro, cuello20flyby, vorobyov20}.
The intruder may also possess a natal disk prior to the flyby. The detection of C in both dust and gas emission indicates its possession of circumstellar material, which may originate from either scenario (Methods). A heavy extinction may be caused by, for example, an edge on circumstellar disk.
Adopting conventional assumptions, its 1.3 mm flux density translates to $\sim28M_\oplus$ of solids in the circumstellar environment.
Future observations at mid- to far-infrared wavelengths are needed to probe the spectral energy distribution of C to further constrain its properties.

Simulations have shown that a prograde flyby may exert gravitational torques and lead to shocks and angular momentum redistribution in the disk, and drive material to smaller radii around the binary\cite{pfalzner03, pfalzner08simulation, vorobyov17, winter18, kuffmeier21}. This may trigger accretion outbursts with the accretion rate jumping up by one order of magnitude shortly after pericenter passage\cite{cuello19hydro}. 
This appears consistent with the double burster nature of Z~CMa.
Isolated protostars with masses from 0.1 -- 40 $M_\odot$ spend on average $1-10\%$ of the time in self-induced outbursts\cite{dunham12, meyer19}. Thus, the chance of witnessing a binary in this mass range undergoing accretion outbursts simultaneously is expected to be $\lesssim1\%$ if the two outbursts are independent of each other. Given the small number of protostar binaries that have been monitored with sufficient depth and temporal coverage to enable detections of outbursts, the double burster nature of Z~CMa implies that whatever has triggered the outbursts in one may have triggered those in the other. This is plausible if both are facilitated by the flyby. 

While our multi-wavelength observations of the Z~CMa system lend support for a possible recent flyby event, alternative scenarios accounting for the observations exist. For example, it is possible that C is a self-gravitating clump previously formed in the dense environment at smaller radii that has subsequently been ejected due to dynamical interactions with the stars and/or other clumps\cite{vorobyov16}. Such clumps may form by fragmentation of gravitationally unstable disks\cite{kratter10, basu12}. In such an event, the ejecta drives a bow shock, which may appear to be filamentary streamers in observations\cite{vorobyov20}. The binary nature of Z~CMa, which promotes dynamical scattering, encourages this scenario. However, the bow shock behind the ejecta typically has two edges with roughly equal strengths (and both of them are expected to be within the disk plane). In addition, the resulting streamers are more prominent closer to the ejecta\cite{vorobyov20}, instead of the other way around as observed. We thus conclude that the ejecta scenario is less likely. Recent observations have shown multiple filamentary structures as infall and accretion streamers connecting a cloud to the disk in a few systems\cite{grady99, fukagawa04, pineda20,ginski21, huang21}. However unlike in those cases, Z~CMa's streamer is not known to be connected to a large-scale cloud, and no other tails similar in strength and size have been found.
Three-dimensional gravo-magnetohydrodynamic simulations have also shown that streamers and clumps can form in turbulent fragmentation in filamentary giant molecular clouds\cite{kuffmeier19, lee19}. While it is possible that the observed streamer and source C form in this manner, the fact that the streamer points toward the source C shows that the flyby scenario is compelling.

Our discovery of C at a special location with respect to the streamer, the unique morphology of the disk in multi-wavelength imaging observations, and the double burster nature of Z~CMa show signposts of a flyby in action around a binary young stellar object. These set the Z~CMa system apart from a handful of other flyby candidate systems (Methods). Flybys during star formation are expected to dramatically affect the disk and the planet formation processes within it. For example, the extremely long period object Sedna\cite{brown04} has prompted the hypothesis that the solar system may have experienced close encounters with intruders in its infancy\cite{schwamb10}. Future high spatial resolution observations at both mm and NIR wavelengths may find more systems like Z~CMa by identifying an intruder and its dynamically induced structures, thus establishing an empirical estimate of the frequency of flybys and better assessing its role in the evolution of protoplanetary disks. In particular, it would be exciting to search for flyby events in outbursting protostar systems, to test the hypothesis that flyby may trigger accretion outbursts. Such studies will also shed light on the formation and evolution of binary and multiple systems\cite{duchene13}. 

\clearpage

\clearpage

\section*{Methods}

{\it In the published Nature Astronomy paper, part of the [ALMA observations] and [JVLA observations] sections, as well as the entire [Flyby occurrence rate estimate] section, are moved to Supplementary Information.}

\noindent {\bf Distance of Z~CMa.} 
The distance to Z~CMa is controversial. The Gaia Early Data Release 3 (EDR3) measured a parallax of 2.035 $\pm$ 0.632 mas\cite{lindegren21}, corresponding to a distance of 491$^{+221}_{-116}$ pc. However, the Renormalised Unit Weight Error (ruwe) parameter is 36.798, a value far exceeding the generally accepted upper limit of 1.4. It is a strong implication for unreliable parallax solution\cite{lindegren21}, possibly caused by the photometric variabilities of this system, or by the imperfect modeling of a disk bearing binary system as a single star.

\noindent
In order to clarify the distance of Z~CMa, we performed a Gaia EDR3-based analysis of Canis Major R1, the cluster in which Z~CMa is considered to be a member\cite{claria74, kaltcheva00, takami18}. We compiled a list of 62 young stars in CMa R1, all with ruwe$\leq$1.4, from the literature \cite{herbig88, ikeda08, fernandes15}. The distribution of their distances exhibits a strong single peak at 1125$\pm$30 pc, with only one star located closer than 650 pc. The area occupied by CMa R1 members on the sky coincides well with the position of Z~CMa. The selected cluster members have a tight distribution in proper motion: pm$_{RA}$= -4.49 $\pm$ 0.2 mas yr$^{-1}$, pm$_{DE}$= +1.59 $\pm$ 0.19 mas yr$^{-1}$. The proper motion of Z~CMa, taken either from Gaia EDR3 or from Ducourant et al.\cite{ducourant05}, matches that of CMa R1 within 3$\sigma$ in right ascension, and is higher by not more than 2.25 mas yr$^{-1}$ in declination. We conclude that Z~CMa fits in the CMa R1 cluster in both coordinates and proper motion, and that there is no hint for additional star formation activity in the foreground of the cluster. Thus, we assume that Z~CMa is a member of Canis  Major R1, and adopt a distance of 1125 pc. 

\noindent {\bf ALMA observations.} 
We performed the ALMA (Project 2016.1.00110.S) observations on 2016 October 21 in the C40-6 array configuration.
The minimum and maximum baseline lengths were 15 and 1800 meters.
The gain calibrator was J0730-1141, and the absolute flux and passband calibrator was J0705+1231.
We adopted the 0.86 Jy absolute flux and the $-$0.64 spectral index for J0705+1231 at the representative frequency 223.165 GHz, which were derived from interpolating the grid survey of ALMA.
The bootstrapped flux of J0730-1141 was 1.1 Jy at 223.091 GHz.
There were six spectral windows, of which the (central frequency, total bandwidth, and frequency channel width) are (230.585 GHz, 234 MHz, 61 kHz), (231.695 GHz, 1875 MHz, 976 kHz), (220.394 GHz, 117 MHz, 61 kHz), (219.556 GHz, 59 MHz, 61 kHz), (219.945 GHz, 59 MHz, 61 kHz), (216.812 GHz, 1875 MHz, 488 kHz), respectively.

We performed the ALMA long baseline (C40-8) configuration observations on 2017 September 07 (Project 2016.1.00110.S).
The minimum and maximum baseline lengths are 168 and 6800 meters.
The gain calibrator was J0654-1053, and the absolute flux and passband calibrator was J0705+1231.
We adopted the 0.84 Jy absolute flux and the $-$0.67 spectral index for J0705+1231 at the representative frequency 223.162 GHz.
The bootstrapped flux of J0654-1053 was 0.12 Jy at 223.088 GHz.
The spectral setups were similar to that of the observations on 2016 October 21, except that the spectral windows at 231.695 GHz and 216.812 GHz were compromised to have 7812 kHz and 976 kHz frequency channel widths to suppress the data rate.

We performed the Atacama Compact Array (ACA) observations on 2017 July 09 (Project 2016.2.00168.S).
The gain, passband, and absolute flux calibrators were J0730-1141, J0522-3627, and J0854+2006, respectively.
We adopted the 4.4 Jy absolute flux and the $-$0.46 spectral index for J0854+2006 at the representative frequency 223.144 GHz.
The bootstrapped flux of J0730-1141 was 1.5 Jy at 223.071 GHz.
The spectral setups were similar to that of the observations on 2016 October 21, except that the spectral window at 231.671 GHz has a smaller frequency channel width of 488 kHz.

We manually calibrated the data using the CASA v5.1.1 software package\cite{mcmullin07}.
The gain calibrator for our ALMA long baseline observations in 2017 was faint.
To yield reasonably high signal-to-noise (S/N) ratios when deriving the gain phase  solutions, we first solved the phase offsets among spectral windows using the passband calibration scan.
After applying the phase offsets solution, we then derived the gain phase solution by combining all spectral windows.
The calibration of the ACA observations and the ALMA 12-m array observations in 2016 followed the standard procedure of ALMA quality assurance (i.e., QA2).
After calibration, we fitted the continuum baseline and subtracted it from the spectral line data, using the CASA task {\tt uvcontsub}.
We jointly phase self-calibrated all observations based on the continuum data, and then applied the phase self-calibration solutions to both the continuum and the spectral line data.

The phase self-calibrated ALMA 12-m array and ACA data were jointly imaged.
The Multi-Frequency Synthesis (MFS) imaging of the continuum data were performed using CASA-{\tt clean}.
To achieve a high intensity dynamic range, we imaged the upper sideband and lower sideband data separately using only one Taylor term.
We then smoothed the upper sideband continuum image to the same angular resolution as the lower sideband continuum image, and then linearly combined them.
The root-mean-square (RMS) noise level were estimated from the difference of the lower sideband and the smoothed upper sideband continuum images, which are 18, 26, and 34 $\mu$Jy\,beam$^{-1}$ in the case of Briggs Robust $=2, 0, -2$ weighting, respectively.
The corresponding synthesized beam sizes with these weightings are \beam=0''.11$\times$0''.095 (P.A.=75$^{\circ}$), \beam=0''.078$\times$0''.046 (P.A.=66$^{\circ}$), and \beam=0''.071$\times$0''.042 (P.A.=68$^{\circ}$).

Spectral line imaging were performed using the Miriad software package\cite{sault95}.
We first imaged the ACA data alone.
After correcting for the ACA primary beam attenuation, we multiplied the ACA images with the primary beam of the ALMA 12m dishes using Miriad-{\tt demos}.
We then re-generated the 12m primary beam attenuated ACA visibilities using Miriad-{\tt uvmodel}.
Finally, we jointly imaged the re-generated ACA visibility and the ALMA visibilities using Miriad-{\tt invert, clean, and restor}, and corrected for the ALMA 12m dish primary beam attenuation after then.
We compared the images of a few channels of $^{13}$CO 2-1 produced this way, with those created with CASA-{\tt clean} and CASA-{\tt tclean} using mosaic mode, either applying {\tt multi-scale clean} or not, and with and without imaging interactively (i.e., box cleaning).
We found that the quality of the images created with our Miriad imaging strategy is not distinguishable from the best we can achieve with CASA-{\tt clean} or CASA-{\tt tclean}.
In fact, without performing interactive imaging, we achieved a better image quality with our Miriad imaging strategy than with CASA.
Jointly (mosaic) imaging ACA and ALMA data in CASA is extremely time consuming, such that interactively imaging the large number of spectral channels of our high spectral resolution observations is not practical for us.
When imaging in CASA but without interactively boxing, the extended features (either real emission structures or sidelobe response) located in between the primary beams of the ACA and the ALMA 12m dish often led to divergence of the {\tt clean}, which by our experience is less predictable when the {\tt multi-scale clean} is activated.
When such features are presented, it is in fact also not trivial to decide how to place {\tt clean} boxes, and that the imaging process became more subjective.
Due to the high angular resolution of our observations, without activating {\tt multi-scale clean}, the convergence of CASA-{\tt clean} and {\tt tclean} also became unrealistically slow for us.
Therefore, we argue that our present spectral line imaging strategy is fair.

In continuum emission, we detect four compact sources. Two of them (A and B) compose the binary. The other two, C and D, are to the southwest and northeast of the binary, respectively. We performed 2D gaussian fitting using the task {\tt imfit} in the image domain to measure their properties, and list them in Extended Data Table~\ref{tab:fluxes}. Gas observations show prominent outflow and cavity structures on large scales, which dominate the signals in both the moment 0 (Extended Data Fig.~\ref{fig:gas_big}) and moment 1 maps (Extended Data Fig.~\ref{fig:gas_mom1_big}). $^{13}$CO and C$^{18}$O do not trace the central binary very well, while the SO line does. All of these continuum sources have spectral line counterparts, so they are likely parts of the same system. The previously detected streamer in scattered light in the south of the binary has counterparts in both continuum and line emission. 

\noindent {\bf JVLA observations.} 
We performed JVLA Ka band (29-37 GHz) standard continuum mode observations toward Z\,CMa in the A array configurations on 2016 October 07 and 10 (project code: 16B-080). 
We took full RR, RL, LR, and LL correlator products. 
These observations had an overall duration of 6 hours, with $\sim$2.7 hours of integration on the target source. 
The observations on October 07 and 10 covered the parallactic angle ranges of 333$^{\circ}$$-$14$^{\circ}$ and 343$^{\circ}$$-$24$^{\circ}$, respectively.
After initial data flagging, 27 and 25 antennas were available for the respective of these two tracks of observations. 
The projected baseline lengths covered by these observations are in the range of $\sim$54-3579 $k\lambda$), which yielded a \beam=0$''.$070$\times$0$''.$049 (P.A.=2.5$^{\circ}$) synthesized beam (Briggs Robust $=0$ weighting), and a maximum detectable angular scale of $\sim$1$''.$9.
We used the 3-bit sampler and configured the backend to have an 8 GHz simultaneous bandwidth coverage by 64 consecutive spectral windows, which were centered on 33 GHz sky frequency. 
The pointing and phase referencing centers for our target source is on R.A. = 07$^{\mbox{\scriptsize{h}}}$03$^{\mbox{\scriptsize{m}}}$43$^{\mbox{\scriptsize{s}}}$.164 (J2000), decl. =  $-$11$^{\circ}$33$'$06$''.$200 (J2000). 
Antenna pointing was calibrated approximately every hour.
The complex gain calibrator J0730-1141 was observed for 40 s every 102 s to calibrate atmospheric and instrumental gain phase and amplitude fluctuation. 
We observed the bright quasar 3C147 for passband and absolute flux calibrations. 

We manually followed the standard data calibration strategy using the CASA package, release 5.3.0. 
After implementing the antenna position corrections, weather information, gain-elevation curve, and opacity model, we bootstrapped delay fitting and passband calibrations, and then performed complex gain calibration. 
We applied the absolute flux reference to our complex gain solutions, and then applied all derived solution tables to the target source. 
We have performed gain phase self-calibration to remove the residual phase and delay errors.

We generated images using the CASA task {\tt clean}. 
The image size is 6000 pixels in each dimension, and the pixel size is 0$''.$01. 
To maximize sensitivity, we performed multi-frequency {\tt clean} jointly for all observed spectral windows, which yielded a $\sim$33 GHz averaged frequency and a Stokes I RMS noise level of $\sim$5.2 $\mu$Jy\,beam$^{-1}$. The ALMA components A, B, and C are detected at 9 mm. We perform 2D gaussian fitting in the image domain using the CASA task {\tt imfit} to measure their properties, and list them in the Extended Data Table~\ref{tab:fluxes}. The ALMA source D was not detected. 

\noindent {\bf Dust mass estimates.} We adopt fluxes in continuum maps imaged with Robust $=0$ for dust mass estimates. Assuming that the circumstellar material is optically thin at 1.3 mm, we estimate its dust mass based on the integrated flux density as
\begin{equation}
M_{\rm dust}=\frac{f_\nu d^2}{B_\nu(T_{\rm dust}) \ \kappa_\nu},
\label{eq:mdust}
\end{equation}
where $d$ is the distance to the source, $f_\nu$ is the measured flux density, $B_\nu(T_{\rm dust})$ is Planck function at the dust temperature $T_{\rm dust}$, assumed to be 20 K, and $\kappa_\nu$ is the dust opacity $\kappa$, assumed to be 2.2 cm$^2$ g$^{-1}$ at 224 GHz\cite{beckwith90}. The dust mass around C and D are estimated to be 28$\pm$1 $M_\oplus$ and 22$\pm$1 $M_\oplus$, respectively (the uncertainty only reflects the uncertainty in the flux measurement). For the streamer, 
given its measured total flux density in 1.3 mm continuum emission of $3\pm0.6$ mJy, we estimate 
$M_{\rm dust}=119\pm24M_\oplus$.
As a reference, the 3-$\sigma$ detection limit in our 1.3 mm continuum emission is 78 $\mu$Jy, corresponding to a dust mass of 3 $M_\oplus$. We note that the derived dust mass has large uncertainties due to the uncertainties in the dust temperature, opacity, and whether the emission is optically thin or thick.

\noindent {\bf Hydrodynamical and Radiative Transfer Simulations.} We model the hydrodynamical evolution of the disk around Z~CMa using the Smoothed Particle Hydrodynamics (SPH) {\sc Phantom} code\cite{price18phantom}. The disk is modeled using $10^6$ gas SPH particles and assuming a total disk mass of $0.3~M_\odot$ (thus a disk-to-star mass ratio of 5\%). The initial disk surface density follows a power-law profile $\Sigma \propto R^{-1}$ and we assume that the disk is vertically isothermal. We adopt a mean Shakura--Sunyaev disk viscosity $\alpha_{\mathrm SS} \approx 0.005$ by setting a fixed artificial viscosity parameter $\alpha_{\mathrm AV}$ = 0.25 and using the `disk viscosity' flag of {\sc Phantom}\cite{lodato10}. The initial disk inner and outer radii are $R_\mathrm{in}=120$~au and $R_\mathrm{out}=840$~au. 
$R_\mathrm{in}$ is not critical in this study since the main focus is on the arm morphology at large distances. The measured accretion rate onto the primary strongly depends on the choice of $R_\mathrm{in}$ and it is expected to increase during the encounter, especially for prograde inclined flybys\cite{cuello19hydro}. 
During the tidal encounter, material in the disk is spread to $\sim2500$~au. 

We model component C as the intruder. We consider D as a low mass clump that does not have a significant impact on the disk dynamics, thus ignored in the simulations. As mentioned in the main text, D may be a fragment that has recently formed from gravitoturbulent fragmentation in the disk or the surrounding cloud, in which scenario it is only visible at mm but not cm wavelength due to its youth and associated insufficient dust growth. For prograde flybys and companion-disk interaction in general\cite{dong16armviewing}, more prominent spiral arms form interior to the orbit of the perturber. The fact that D appears to be closer to the primaries than the streamer suggests that it is unlikely the driver of the streamer. Moreover, if D was a massive perturber deeply penetrating the disk, it would be expected to leave strong signatures in the disk, which is not observed in our multi-wavelength observations. In contrast, C is a much more robust candidate, since it leaves the expected dynamical signatures in the disk.

We place the perturber on a parabolic prograde orbit at an angle of $\beta=45^{\rm o}$ with respect to the disk plane as shown in the sketch in Extended Data Fig.~\ref{fig:sketch}. The pericentre of the orbit is fixed along the line of nodes and at a distance of $r_{\rm peri}=3000$~au from the disk-hosting star. The perturber is initially placed at ten times $r_{\rm peri}$. The masses of the host star and the perturber are set to 6 M$_\odot$ and 1.8 M$_\odot$, which give a mass ratio $q=0.3$. Previous simulations\cite{cuello19hydro, pfalzner03} have shown that as the mass ratio increases, the flyby event produces more prominent streamers, while the intruder rips off a larger fraction of the circumprimary disk, leaving a more compact disk afterward. The exact radial extent and morphology of the streamer depend on the mass ratio $q$, pericentre distance $r_{\rm peri}$, initial disk size, and orbital inclination $\beta$. The mass of C is poorly constrained from the observations. Based on a limited number of tests we find that a minimum mass ratio of $\sim0.1$ is needed in order to roughly reproduce the streamer with a companion at the current distance. We model both the star and the perturber as sink particles that interact with gas particles via gravity and accretion\cite{bate95}. Given the large spatial scales considered and for computational costs, we set the sink particle accretion radius to 10~au on both stars. 

The flyby setup is publicly available in the {\sc Phantom} repository. Here we consider a non-penetrating prograde flyby in order to trigger prominent spirals without capturing material around the perturber. 
The circumprimary disk is initially in the $x-y$ plane. 
In order to simultaneously match the appearance of the disk in scattered light and the orientation of C, we rotate the system by $105^\circ$, $30^\circ$, and $180^\circ$ around the $z$-, $y$-, and $x-$axis (in clockwise sense and in that order). The line of sight is in the $z$ direction (axis are shown in red in Fig.~\ref{fig:simulation} and Extended Data Fig. \ref{fig:sketch}). 

We use the radiative transfer code {\sc mcfost}\cite{pinte06, pinte09} to generate synthetic polarized scattered light images. We build a Voronoi tesselation where each cell is associated to one SPH particle. The disk is passively heated by the two stellar sources: the host, and the intruder; however, the flyby star is turned off in simulating the scattered light image to mimic extinction by its circumstellar material (see below). We use isochrones\cite{siess00, allard12} at 3~Myr for the two sink particles of 6 and 1.8~$M_\odot$, resulting in effective temperatures of 18400 and 4800 K and radii of 3.00 and 2.36 $R_\odot$, respectively. We used $10^8$ packets to compute the dust temperature structure and sample the specific intensity before generating the polarized scattered light images at 1.6\,$\mu$m via a ray-tracing algorithm. The dust grain population is assumed to follow a power-law $\mathrm{d}n(a) \propto a^{-3.5}\mathrm{d}a$ between 0.03 micron and 1 micron, and the total dust mass is set to 0.01 times the gas mass from the SPH simulation. 
While grains are known to be able to grow to larger sizes in early stage disks\cite{vorobyov18}, sub-micron-sized grains dominate the scattering of starlight. The presence of larger grains, not included in our simulations, is not expected to affect scattered light observations.
The dust properties are calculated using the Mie theory assuming astrosilicate composition\cite{weingartner01}. The resulting synthetic observation in the H-band are shown in Fig.~\ref{fig:simulation}.

A prograde perturber may cause a strong gravitational pull able to excite the eccentricity of the material in the disk. This translates into enhanced stellar accretion and outbursts. The time delay between the pericenter passage and the peak of the accretion outbursts, as well as their duration, depend on a number of parameters including the intruder-flyby mass ratio, the pericenter distance, the angle of the pericenter and the inclination of the intruder's orbit relative to the circum-primary disk plane, and the size of the circum-primary disk. Cuello et al.\cite{cuello19hydro} carried out a sample of tests and found that for prograde flyby events the delay between the pericenter passage and the peak of the outbursts is on the order of 10\% of the Keplerian orbital period at the pericenter, and the outbursts temporally coincide with prominent spiral arms in the disk (they both fade away with time). Meanwhile the outbursts can last for at least thousands of years. These are broadly consistent with our model. We caution that in current simulations\cite{cuello19hydro} the stellar accretion is modeled as material crossing the inner boundary of the disk at 10 AU. Future investigations are needed to model the disk dynamics in the inner disk to better quantify the accretion outburst timescales.

Our model demonstrates that a flyby event can produce one prominent arm extending beyond the outer edge of the circumprimary disk and pointing to the intruder. We note that the model is not designed to perfectly reproduce the observations, and no extensive parameter space survey has been carried out to find a best fit model or to constrain the parameters in the flyby. Despite a match in the overall shape, there are a number of discrepancies between the model and the observations. For example, the streamer extends to a larger distance and continuously fades outward in the synthetic NIR image, while the observed streamer bends at its end and ends more sharply. Moreover, the synthetic image shows another spiral arm in the north side of the disk  interior to 1$''$, which is not seen in observed scattered light (although there are hints of a secondary arm at this distance in the ALMA dust and gas emission maps; Figs.~\ref{fig:basic} and \ref{fig:gas_small}). The observed streamer is also more than one order of magnitude brighter than that in the model. These discrepancies are likely the result of the model having a different orbit of the intruder, a different initial disk size, a different viewing orientation, and/or different scattering properties of the dust from the Z~CMa system. In addition, the signal-to-noise ratio interior to 1$''$ in the Subaru observation may not be high enough to reveal structures.

\noindent {\bf Keck/NIRC2 archive data.} The ALMA source~C is located outside the field-of-view (FoV) in Subaru/HiCIAO observations (Fig.~\ref{fig:basic}a). In Extended data Fig.~\ref{fig:keck} we present $H$-band Keck/NIRC2 observations taken on 2005 Oct 21 (Keck program ID: U14N2), which have a FoV large enough to cover C (10$''$~$\times$~10$''$). The Keck data were taken with a coronagraph. The total exposure time is 10~seconds. We conducted the standard calibration procedure: dark subtraction, flat fielding, and hot/bad pixel correction. No point sources are detected at the position of source~C. According to the NIRC2 observer's manual, the zero-point magnitude at the $H$-band is 25.4~mag. The background noise  including photon noise of the central star and the sky noise at the position of the point source~C is 23.7~mag, and thus, the 5~$\sigma$ detection limit of the point source~C at the $H$ band is 21.9~mag in apparent magnitude. 

\noindent {\bf Circumstellar extinction of component C.} Our experiments show that a flyby object with $q\gtrsim0.1$ is needed to induce the observed streamer, corresponding to $M_C\gtrsim0.6M_\odot$. A 2 Myr, a $0.6M_\odot$ YSO has a photospheric luminosity of 13.3 mag at 1125 pc in apparent magnitude at $H$-band\cite{baraffe15}. Thus, an extinction of $A_H>8$ mag is needed to render component C undetected in the Keck observations. 
A flyby object on a grazing or penetrating orbit may capture material from the primary disk\cite{forgan10, cuello19hydro, cuello20flyby, vorobyov20}. It may also possess circumstellar material prior to the flyby.
The estimated dust mass in the circumstellar material around C is 28$\pm1M_\oplus$ based on its 1.3 mm flux, which may provide sufficiently large extinction. For example, if one Lunar mass of interstellar medium dust with $\kappa=$5300 cm$^2$ g$^{-1}$ at $H$-band\cite{takami13} are distributed spherically symmetrically with a radial density profile $\rho\propto r^{-1.5}$ between 1 to 10 au, the resulting extinction at $A_H$ reaches 10 mag. 
Alternatively, if C acquires an edge on disk after the flyby, the stellar component can be completely obscured\cite{terada07} too. Modeling the specific circumstellar environment around C after the flyby is difficult, due to our ignorance of its state prior to the flyby.

We note that Class 0 YSOs, protostars embedded in a substantial natal envelope, have spectral energy distributions peaking around or longward of $100\mu$m\cite{frost19}. Such objects are often detectable only at far-infrared and millimeter wavelengths\cite{gerin17}. However, source C is unlikely to belong to this class. Its gas emission is confined to a region ~200 au in size (Fig.~\ref{fig:gas_small}), inconsistent with the thousands of au size of the envelope of Class 0 objects\cite{gaudel20}. In addition, the envelope would likely be distorted and stretched due to the gravitational interaction with the primary, which is not seen.
Class 0 objects usually have $M_\star < M_{\rm envelope}$\cite{floresrivera21}. The observed 1.3 mm flux from C corresponds to a total gas+dust mass of 0.01 $M_\odot$, assuming a gas-to-dust mass ratio of 100, which is two orders of magnitude smaller than the estimated mass of component C.

\noindent {\bf Comparisons of the Z~CMa system with other flyby candidate systems.} A handful of systems with disks undergoing possible stellar flybys have been proposed before, such as RW~Aur\cite{cabrit06,dai15}, AS~205\cite{kurtovic18}, and UX~Tau\cite{menard20, zapata20}. However, the unbound nature of the perturber in them does not have strong observational support and remains to be confirmed. For example, latest observations suggest that the two components in RW~Aur form a bound binary, as the disk morphology is inconsistent with a one-off encounter\cite{rodriguez18}. 
In the bound companion case the circumprimary disk is expected to be severely truncated and the spirals tightly wound and symmetric\cite{dong16hd100453,wagner18}. The radial extent and the shape of the spiral and the disk in Z~CMa strongly favor an unbound perturber. In addition, multiple encounters with a bound companion drive a pair of symmetric arms even if the orbit of the companion is misaligned from the circumprimary disk\cite{vanderplas19, rosotti20}. This is not seen in Z~CMa but seen in other proposed flyby systems. 
The streamer in Z~CMa is convincingly detected in all three tracers (scattered light, and gas and dust emission), which is not the case in other systems.
Moreover, the more compact streamer in mm dust continuum than in scattered light, as well as the burster nature of the hosts in Z~CMa, support a recent flyby as shown by simulations\cite{cuello19hydro, cuello20flyby}.

\noindent {\bf Flyby occurrence rate estimate.} In young clusters with a stellar mass density $\rho_{\rm cluster}$ of a few 100s$-$1000 $M_\odot$ pc$^{-3}$, one can estimate the average time required for a particular star to experience a flyby by another star passing within a pericenter distance $r_{\rm peri}$ as\cite{davies11, pfalzner13}:
\begin{equation}
\tau \sim 33\ {\rm Myr} \left(\frac{50\ M_\odot\ {\rm pc}^{-3}}{\rho_{\rm cluster}}\right) \left(\frac{v_\infty}{1\ {\rm km\ s}^{-1}}\right) \left(\frac{1000\ {\rm au}}{r_{\rm peri}}\right) \left(\frac{M_\odot}{M_{\rm t}}\right)
\label{eq:tau}
\end{equation}
where $v_\infty$ is the mean relative speed of the stars in the cluster at infinity, and $M_{\rm t}$ is the total mass of the stars involved in the encounter. The last term represents the effect of gravitational focusing, as cross sections are increased as stars are deflected towards each other due to mutual gravitational attraction. Large disks around massive stars are most susceptible to close encounters. 

For $r_{\rm peri}\sim3000$ au, $M_{\rm t}\sim10M_\odot$, $v_\infty\sim1$ km s$^{-1}$ (typical velocity dispersions in low mass star forming clusters\cite{cottaar15, foster15}), and $\rho_{\rm cluster}\sim200M_\odot\ {\rm pc}^{-3}$ (typical for nearby star forming clusters\cite{lada03}), the average encounter time is about 0.3 Myr,
comparable to the age of Z~CMa. 
Other encounter timescale estimates yield similar results\cite{clarke91, forgan10}.
Patchy stellar distribution within dense groups\cite{joncour18} and
mass segregation leading to massive stars, such as Z~CMa, being located in the central region with the highest density could further shrink the encounter timescale. Although the stellar mass density in the environment of Z~CMa is not known accurately, we note that the binary is located close to the center of a star-forming molecular gas filament over $3'$ ($\sim$1 pc) in length\cite{sandell01}, and a relatively high $\rho_{\rm cluster}$ is expected.
Overall, it is reasonable to expect the Z~CMa system to have experienced disk-perturbing flyby events at its young age. Simulations have also shown that\cite{pfalzner08fuor, pfalzner13, bate12, bate18} such encounters are common in the center of young dense clusters, and outburst-inducing encounters are more likely for massive stars.

\subsection{Data availability} Raw ALMA data is publicly available via the ALMA archive \url{https://almascience.eso.org/aq/} under project IDs 
2016.1.00110.S and 2016.2.00168.S. 
Raw JVLA data is publicly available via the JVLA archive \url{https://archive.nrao.edu/archive/advquery.jsp} under project code 16B-080. 
Raw Keck data is publicly available via the KOA Data Access Service http://koa.ipac.caltech.edu/ under the program ID U14N2. Final reduced and calibrated image files are available at \url{https://doi.org/10.6084/m9.figshare.16915327}.

\subsection{Code availability} The {\sc Phantom} code is made available at https://github.com/danieljprice/phantom by Daniel Price. The {\sc mcfost} code is made available at https://github.com/cpinte/mcfost by Christophe Pinte.

\clearpage

\begin{addendum}
  
  \item[Supplementary Information] is available in the online version of
  the paper.

  \item We thank Chul-Hwan Kim, Jeongeun Lee, and Hongping Deng for helpful discussions. R.D. would like to thank Luyi Xu for the support and encouragement in the period of this work. R.D. acknowledges financial support provided by the Natural Sciences and Engineering Research Council of Canada through a Discovery Grant, as well as the Alfred P. Sloan Foundation through a Sloan Research Fellowship. N.C. acknowledges support from the European Union's Horizon 2020 research and innovation programme under the Marie Sk\l{}odowska-Curie grant agreement No 210021. H.B.L. acknowledges support from the Ministry of Science and Technology (MoST) of Taiwan (grant No. 108-2112-M-001-002-MY3). E.I.V. acknowledges support from the Russian Fund for Fundamental Research,  Russian-Taiwanese  project  19-52-52011. Y.H. is supported by the Jet Propulsion Laboratory, California Institute of Technology, under a contract with the National Aeronautics and Space Administration. A.K. and L.C. acknowledge support from the European Research Council (ERC) under the European Union's Horizon 2020 research and innovation programme under grant agreement No 716155 (SACCRED). L.C. is supported by the Hungarian OTKA grant K132406. M. Takami is supported by the Ministry of Science and Technology (MoST) of Taiwan (grant No. 106-2119-M-001-026-MY3, 109-2112-M-001-019). H.B.L. and M.T. and are supported by MoST of Taiwan 108-2923-M-001-006-MY3 for the Taiwanese-Russian collaboration project. The Geryon cluster at the Centro de Astro-Ingenieria UC was extensively used for the calculations performed in this paper. BASAL CATA PFB-06, the Anillo ACT-86, FONDEQUIP AIC-57, and QUIMAL 130008 provided funding for several improvements to the Geryon cluster. This paper makes use of the following ALMA data: ADS/JAO.ALMA\#2016.1.00110.S and \#2016.2.00168.S. ALMA is a partnership of ESO (representing its member states), NSF (USA) and NINS (Japan), together with NRC (Canada), MOST and ASIAA (Taiwan), and KASI (Republic of Korea), in cooperation with the Republic of Chile. The Joint ALMA Observatory is operated by ESO, auI/NRAO and NAOJ. The National Radio Astronomy Observatory is a facility of the National Science Foundation operated under cooperative agreement by Associated Universities, Inc. 
 
  \item[Author Contributions] R.D. led the ALMA proposals. H.L. led the JVLA proposal and processed the ALMA and JVLA data. N.C. and C.P. performed the hydrodynamics and radiative transfer simulations. R.D., H.L, N.C., and C.P. wrote the manuscript. All co-authors provided input to the observational proposals and/or the manuscript.

  \item[Competing Interests] The authors declare that they have no competing financial interests.
 
  \item[Correspondence] Correspondence and requests for materials should be addressed to: 

R.D.~(email: rbdong@uvic.ca), 

H.L.~(email: hyliu@asiaa.sinica.edu.tw), 

N.C.~(email: nicolas.cuello@univ-grenoble-alpes.fr),

C.P.~(email: christophe.pinte@monash.edu).
 
\end{addendum}

\clearpage

\section*{Extended Data (Figures and tables in the Methods Section)}
\renewcommand{\figurename}{\textbf{Extended data figure}}
\renewcommand{\tablename}{\textbf{Extended data table}}
\setcounter{figure}{0}    

\begin{figure}
\includegraphics[width=0.9\textwidth,angle=0]{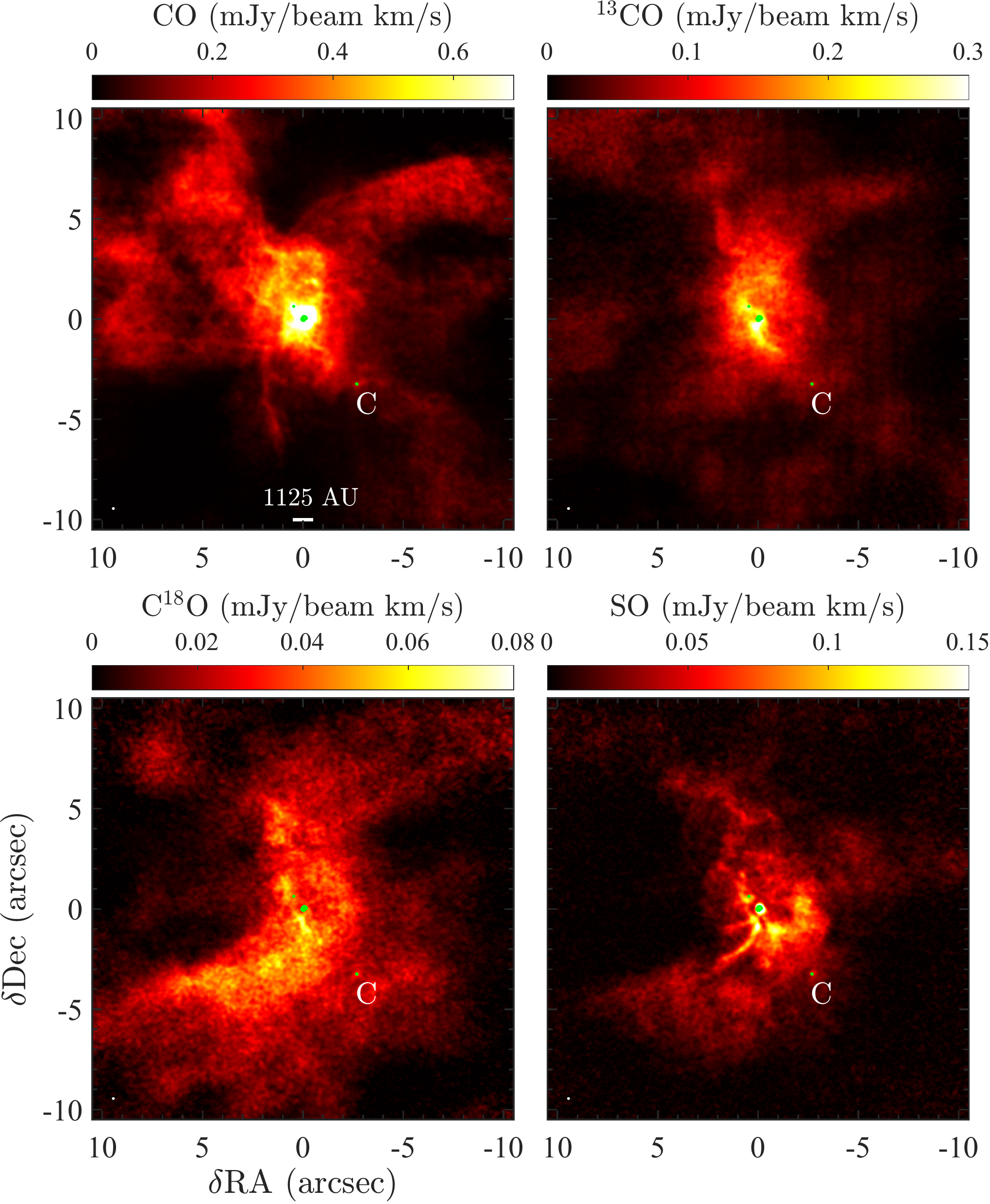}
\caption{{\bf Integrated intensity (moment 0) maps of the $^{12}$CO (2-1), $^{13}$CO (2-1), C$^{18}$O (2-1), and SO $^{3}\Sigma$ 6(5)-5(4) transitions on a large scale.} The beam size ($0.19''\times0.18''$; P.A.=88$^\circ$) is marked at the lower left corner. The Robust $=0$ weighted 224 GHz continuum image (beam size $0''.075\times0''.047$; P.A.=65$^{\circ}$) is shown in green contours at 0.13 and 1.3 mJy beam$^{-1}$ levels (5 and 50 $\times$ the root mean square noises) to highlight the four compact continuum sources. The ALMA/VLA source C is labeled.}
\label{fig:gas_big}
\end{figure}

\begin{figure}
\includegraphics[width=0.9\textwidth,angle=0]{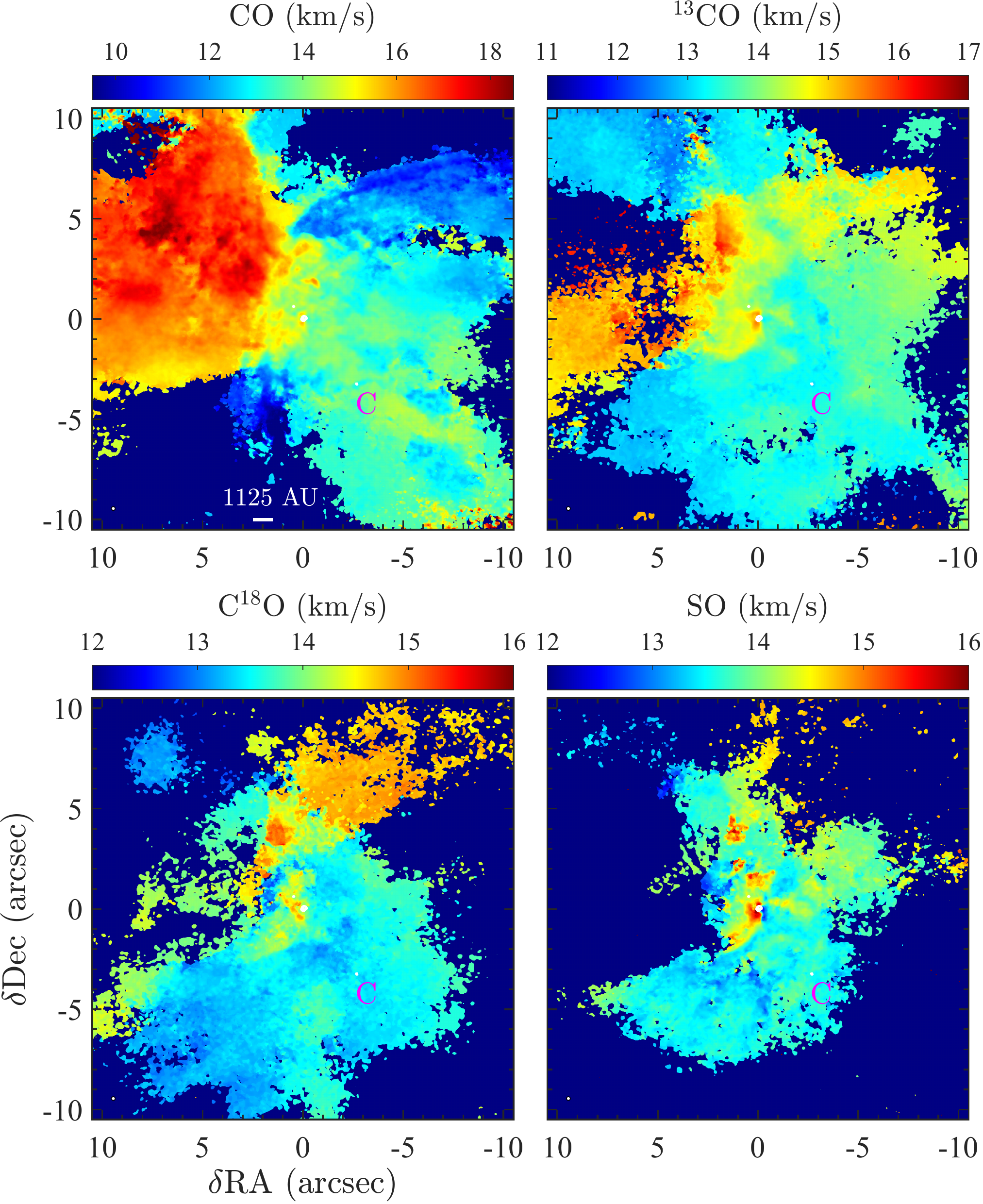}
\caption{{\bf Moment-1 maps (intensity-weighted mean velocity) of the gas observations.} The Robust $=0$ weighted 224 GHz continuum image (beam size $0''.075\times0''.047$; P.A.=65$^{\circ}$) is shown in white contours at 0.13 and 1.3 mJy beam$^{-1}$ levels (5 and 50 $\times$ the root mean square noises). The systematic velocity is $\sim14$ km/s \cite{liljesterom97}. The ALMA/VLA source C is labeled.}
\label{fig:gas_mom1_big}
\end{figure}

\begin{figure}
\includegraphics[width=0.9\textwidth,angle=0]{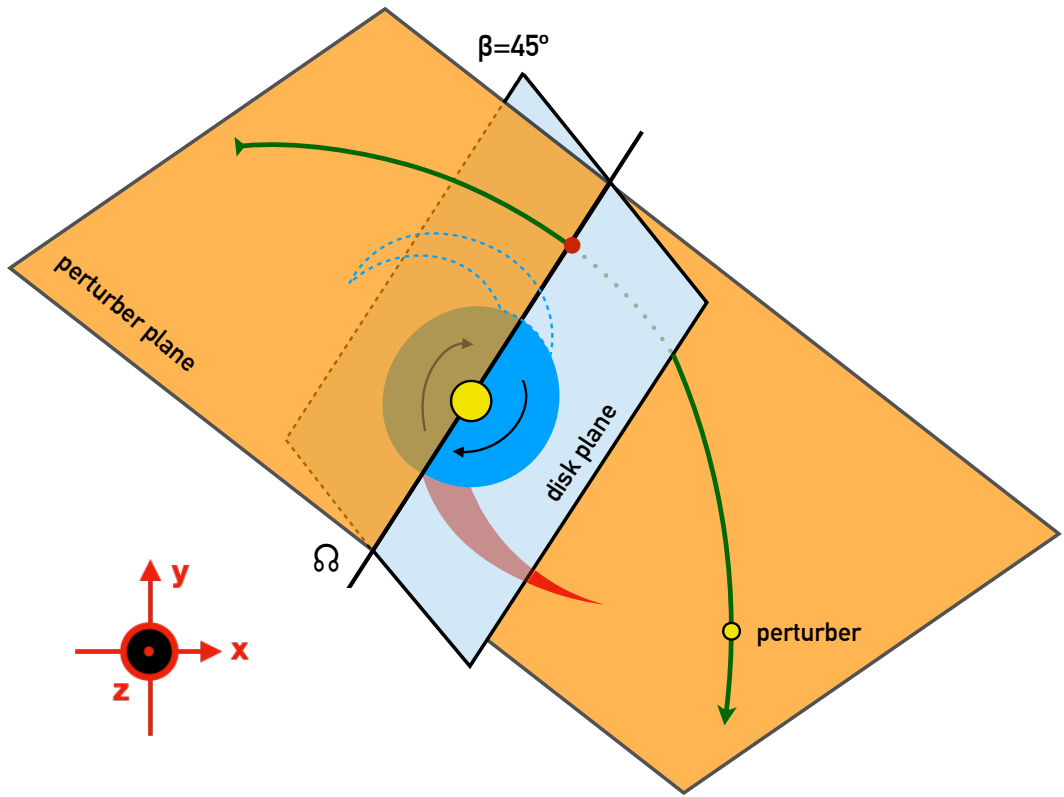}
\caption{{\bf Sketch of the flyby simulation geometry.} The perturber's orbital plane is in orange. The plane of the initial circumprimary disk is in blue. The primary spiral pointing towards the perturber is in solid red and is in the plane of the perturber. The secondary spiral is in dashed red and in the plane of the circumprimary disk. The line of sight is in the $z$ direction. 
}
\label{fig:sketch}
\end{figure}

\begin{figure}
\centering
\includegraphics[width=0.5\textwidth,angle=0]{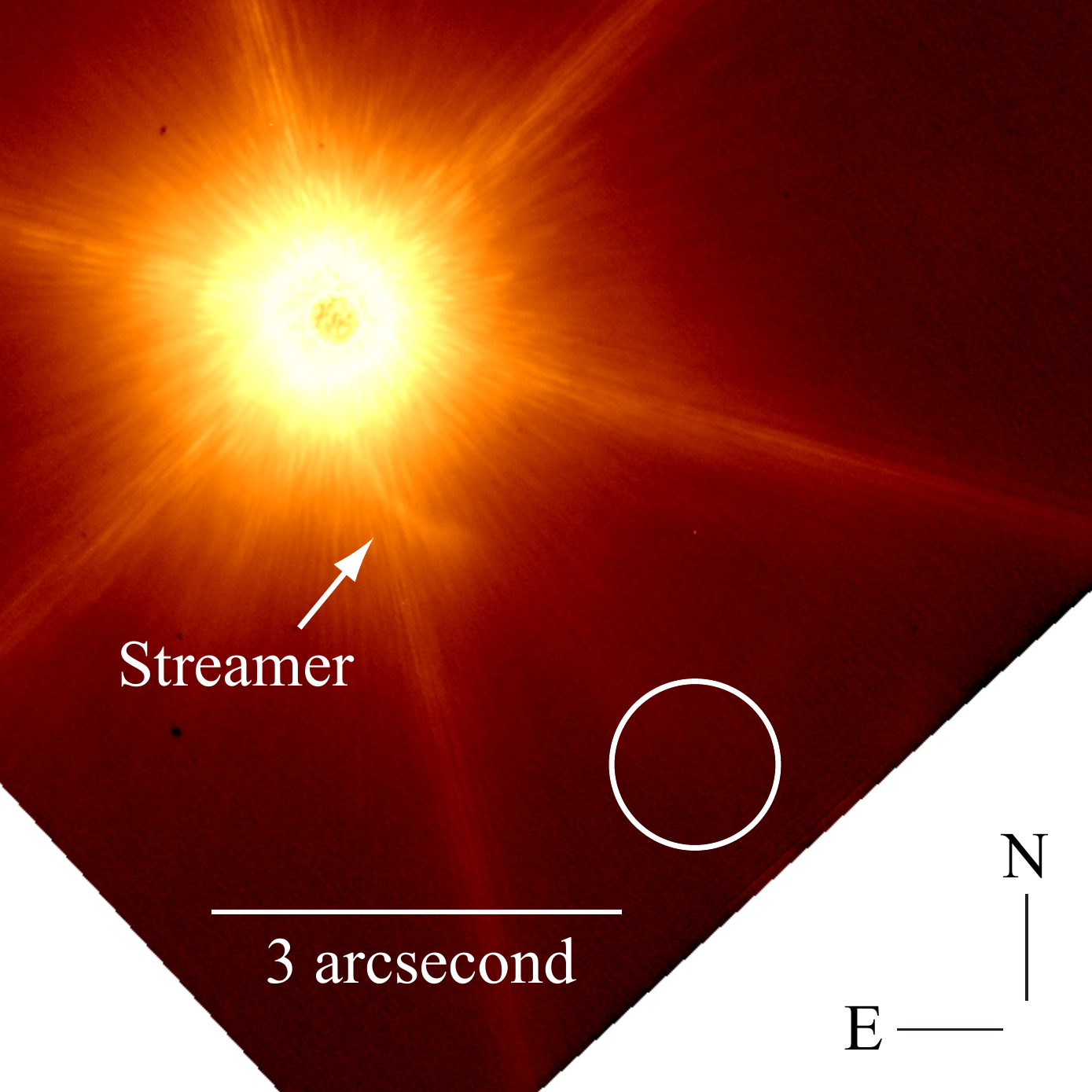}
\caption{{\bf Keck/NIRC2 $H$-band archive data of Z~CMa.} The data were taken in 2005 Oct 21. White circle indicates the position of the point source C.}
\label{fig:keck}
\end{figure}

\begin{figure}
\includegraphics[width=\textwidth,angle=0]{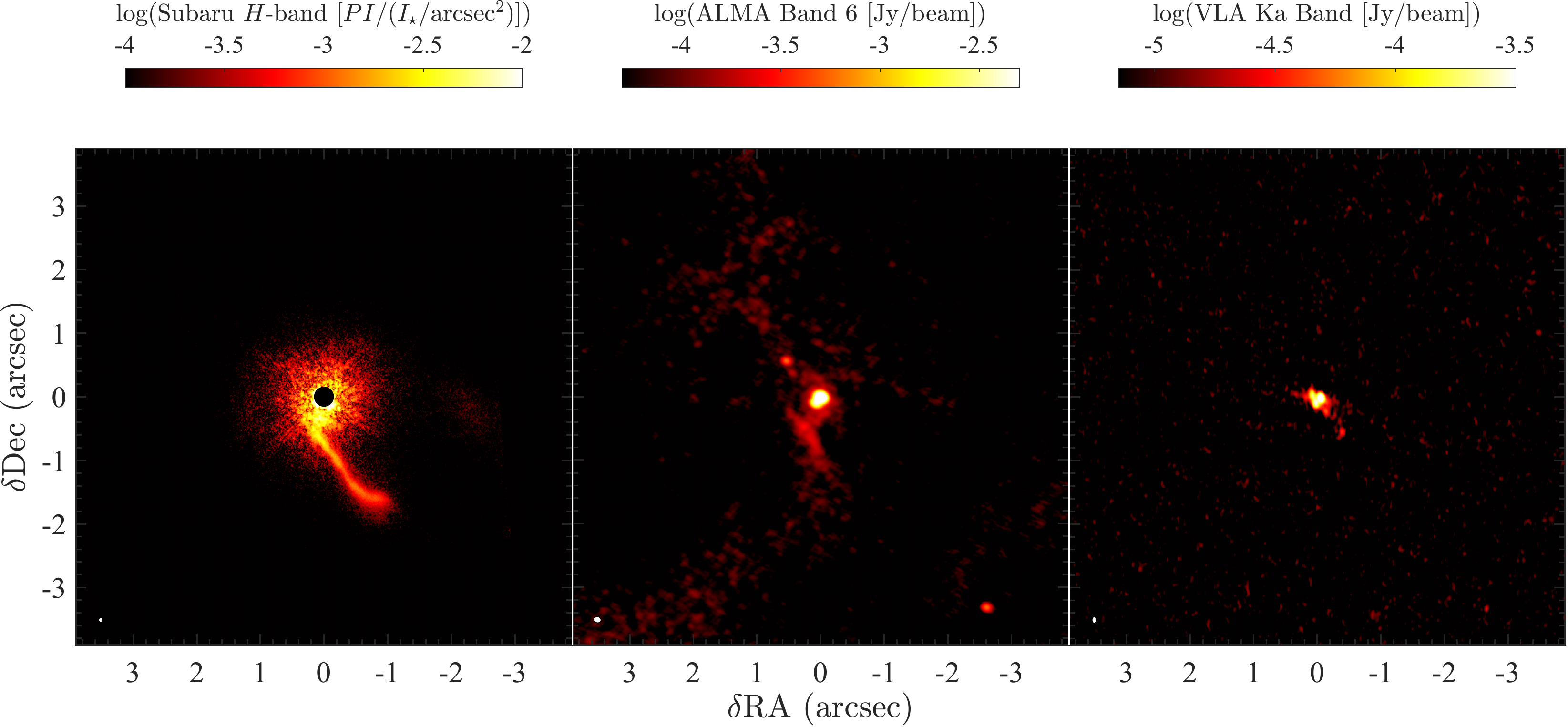}
\caption{The same as Figure~1 in the main text, but without annotations and insets.}
\label{fig:basic_noannotation}
\end{figure}

\begin{table}
  \caption{Gaussian fitting of point-like components detected around Z\,CMa~$^{a}$}\label{tab:fluxes}
  {\footnotesize
  \begin{tabular}{cccccccc}
  \hline
  \# & R.A. & Decl. & Peak intensity & Gaussian fitted flux & $a$ $^{b}$ & $b$ $^{c}$ & P.A. $^{d}$ \\
   & (J2000) & (J2000) & ($\mu$Jy\,beam$^{-1}$) & ($\mu$Jy) & (mas) & (mas) & (degree) \\
  \hline
  \multicolumn{8}{c}{ALMA 224.247 GHz, Briggs Robust $=0$ weighted (\beam=0$''$.078$\times$0$''$.046; P.A.=66$^{\circ}$)} \\
  \hline
A & 07\rah03\ram43\ras.154 & -11\decd33\decm06\decs.13 & 21000$\pm$410 & 24000$\pm$810 & 23$\pm$6.8 & 22$\pm$9.1 & 82$\pm$86 \\
B & 07\rah03\ram43\ras.159 & -11\decd33\decm06\decs.22 & 4100$\pm$62 & 4700$\pm$120 & 26$\pm$5.8 & 20$\pm$5.9 & 71$\pm$61 \\
C & 07\rah03\ram42\ras.976 & -11\decd33\decm09\decs.43 & 480$\pm$26 & 700$\pm$40 & 43$\pm$8.5 & 35$\pm$7.6 & 51$\pm$84 \\
D & 07\rah03\ram43\ras.191 & -11\decd33\decm05\decs.56 & 560$\pm$26 & 550$\pm$24 & $\cdots$ $^{f}$ & $\cdots$ & $\cdots$ \\
  \hline
  \multicolumn{8}{c}{JVLA 32.998 GHz, Briggs Robust $=0$ weighted (\beam=0$''$.070$\times$0$''$.049; P.A.=2.5$^{\circ}$)} \\
  \hline
A & 07\rah03\ram43\ras.153 & -11\decd33\decm06\decs.14 & 630$\pm$28 & 780$\pm$57 & 37$\pm$7.7 & 18$\pm$13 & 80$\pm$49 \\
B & 07\rah03\ram43\ras.158 & -11\decd33\decm06\decs.22 & 260$\pm$11 & 390$\pm$29 & 59$\pm$9.8 & 33$\pm$12 & 38$\pm$23 \\
C & 07\rah03\ram42\ras.975 & -11\decd33\decm09\decs.46 & 24$\pm$5.2$^{e}$ & 27$\pm$5.2 & $\cdots$ & $\cdots$ & $\cdots$ \\
  \hline
  \end{tabular}
  \par
  \begin{itemize}
    \item[$^{a}$] Gaussian fittings may be confused by extended dust emission. While we treat the Gaussian fitted flux as the total flux and use it in dust mass estimates, the true total flux from each source may be in between the peak intensity and the Gaussian fitted flux.
    \item[$^{b}$] Gaussian deconvolved major axis FWHM.
    \item[$^{c}$] Gaussian deconvolved minor axis FWHM.
    \item[$^{d}$] Gaussian deconvolved position angle.
    \item[$^{e}$] Only detected from the Robust $=2$ weighted image, with \beam=100$\times$66 mas (P.A.=19$^{\circ}$).
    \item[$^{f}$] Gaussian fitting results indicate that the source is consistent with a $\delta$ function with non-detectable size.).
  \end{itemize}
  }
\end{table}

\end{document}